\documentclass[9pt,twocolumn,twoside]{osajnl}

\usepackage{mwe}
\usepackage{subfig}
\usepackage{xcolor}
\renewcommand\fbox{\fcolorbox{white}{white}}
\usepackage{graphicx}

\journal{ol} 

\setboolean{shortarticle}{true}

\title{Broadband and robust adiabatic second harmonic generation by temperature gradient in birefringently phase matched lithium triborate crystal}

\author[1,*]{Eyal Rozenberg}
\author[1]{Ady Arie}

\affil[1]{School of Electrical Engineering, Tel Aviv University, Tel Aviv, Israel 69978}
\affil[*]{Corresponding author: eyal.rozenberg1@gmail.com}

\begin{abstract}
Phase-matched nonlinear processes exhibit a tradeoff between the conversion efficiency and the acceptance bandwidth. Adiabatic nonlinear processes, in which the phase mismatch varies slowly along the interaction length, enable us to overcome this tradeoff, allowing an efficient frequency conversion with broad spectral and thermal bandwidths. Until now, the variation in the phase mismatch condition was mainly based on quasi-phase matching in ferroelectric crystals. However, this solution is limited to low power sources. Here, instead, we study the adiabatic second harmonic in birefringently phase-matched lithium triborate crystal, enabling us to handle much higher power levels. The variation in the phase mismatch is achieved by inducing a temperature gradient along the crystal. By using a 50 mm long crystal, the adiabatic process provided a temperature bandwidth of 18$^{\circ}$C, 5.4 times wider than what is achieved when the same crystal is held at the fixed phasematching temperature. The conversion efficiency exceeded 60\% for a 0.9 millijoule pump pulse.
\end{abstract}

\setboolean{displaycopyright}{false}
\begin{document}
\maketitle

The efficiency of nonlinear frequency conversion processes relies on the fulfillment of the phase matching condition \cite{PhysRev.127.1918}. The standard way for achieving it is by maintaining perfect and constant phase matching between the interacting waves, either by relying on the birefringence of the nonlinear crystal or by providing an additional crystal momentum through periodic modulation of the nonlinear crystal in a quasi-phase-matched process. However, this standard solution is sensitive to variations in the pump’s wavelength, its propagation direction, and the crystal temperature, and provides low efficiency for wideband sources. In the last decade, a different method was proposed and studied – adiabatic frequency conversion \cite{PhysRevA.78.063821,doi:10.1002/lpor.201300107} – in which the process starts with a large negative phase mismatch, this mismatch varies slowly along the interaction and the process ends with a large positive mismatch. Adiabatic frequency conversion is a robust process, enabling to achieve high conversion efficiency with large thermal, angular and spectral bandwidths.

The research on adiabatic frequency conversion processes started in the case of strong and nearly undepleted pump \cite{PhysRevA.78.063821,doi:10.1002/lpor.201300107,Suchowski2009RobustAS}, in which the pump wave can be assumed to be constant and the system can be modeled as a two-level system; where the two remaining waves are coupled by the nonlinearity of the crystal. This process is attractive for broadband chirped pulse amplification \cite{Heese:10, Heese:12} and can provide nearly 100\% conversion efficiency, spanning over more than one octave at optical frequencies \cite{Moses:12, Krogen}.  The theory of adiabatic nonlinear processes can be extended to the fully nonlinear case \cite{1993JETPL..57..790B,Baranova_1995,Yaakobi:13,Yaakobi:13_2,Porat:13,Leshem:16,Dahan_2017}, in which all the interacting waves can be depleted. A criterion of the adiabaticity for fully nonlinear interaction was also defined \cite{Porat:13}. These were experimentally demonstrated in type I and type II second harmonic generation (SHG) processes with nanosecond pulses \cite{Leshem:16}, and recently also studied with ultrafast pulses \cite{Dahan_2017}. In both cases, efficient and wideband conversion was achieved; however, the experimental realizations relied on a quasi-phase-matched interaction in aperiodically-poled KTiOPO$_4$, where the variation in phase mismatch was achieved by varying the poling period along the crystal. Unfortunately, the ferroelectric crystals (e.g. KTiOPO$_4$, LiNbO$_3$, LiTaO$_3$) that are used for quasi-phase matched interactions have limited damage threshold ($\sim$500 MW/cm$^2$ for 10 nsec pulses @1064 nm \cite{nikogos}) and small aperture (typical crystal thickness of $\sim$1 mm), thus limiting the power level that can be used for adiabatic frequency conversion.

\begin{figure}[htbp]
\centering
\fbox{\includegraphics[width=\linewidth]{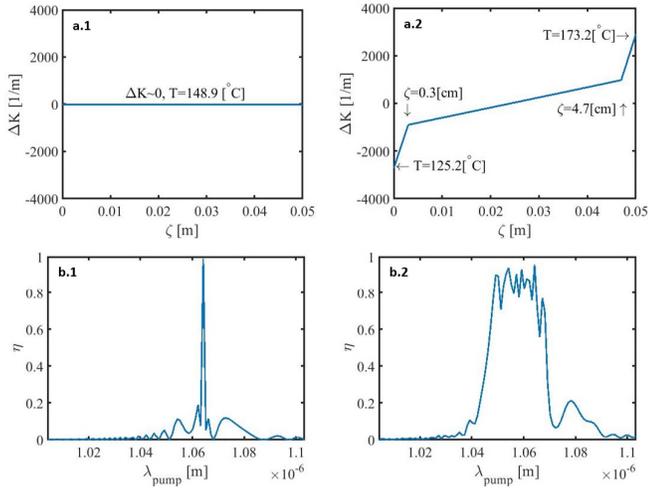}}
\caption{(a) wave vector mismatch  $\Delta$k as a function of interaction length $\zeta$[m]. (a.1) constant value, (a.2) adiabatic variation and (b) the corresponding conversion efficiency as a function of pump wavelength. Simulated on a 5 cm LBO crystal properties with pump intensity of 1.1GW/cm$^2$.}
\label{fig:picture1}
\end{figure}

To extend the method of adiabatic conversion for high-power and high-intensity sources, it is required to develop a method that relies on crystals that utilize birefringent phase matching and can withstand high power levels. Specifically, lithium triborate (LBO) provides high damage threshold ($\sim$45GW/cm$^2$ for 1.1 nsec pulses @1064 nm \cite{nikogos}) and can be obtained with very large apertures.  A type-I noncritical phase matching of LBO at 1064 nm is obtained at a crystal temperature of 149$^{\circ}$C \cite{362711, doi:10.1063/1.360499}, but the spectral and thermal bandwidths are quite narrow \cite{161322}, $\delta\lambda$<1 nm and $\delta$T<1$^{\circ}$C for a 5 cm long crystal. The spectral response for the case of a fixed temperature (Fig. \ref{fig:picture1}a.1) is shown in Fig. \ref{fig:picture1}b.1. In order to achieve wider bandwidths, we can utilize the concept of adiabatic conversion, which can be realized by a thermal gradient: At the entrance to the crystal, the temperature will be below the phase matching temperature, hence the process is not phase matched and the two harmonic waves (fundamental and second harmonic) are uncoupled. The temperature will increase slowly (adiabatically) along the crystal, passing the phase matching temperature at the middle of the crystal, and rising towards the exit of the crystal to a temperature which is much higher, hence decoupling again the two harmonic waves. An example of the temperature profile and corresponding spectral response for an adiabatic second harmonic process are shown in Fig. \ref{fig:picture1}a.2 and Fig. \ref{fig:picture1}b.2. This concept was first suggested by Yaakobi et al \cite{Yaakobi:14} for the case of SHG in the context of auto-resonant processes \cite{Yaakobi:13,Yaakobi:13_2} and recently also measured by a thermal gradient in MgO:LiNbO$_3$ \cite{Dimova_2018}, but these reports included only the conversion efficiency vs the pump power, without a detailed study of the acceptance bandwidth and the required thermal profile. More recently, broadband adiabatic optical parametric amplification was reported using a 4-point temperature profile \cite{Markov:18}. At present, there isn't still any theoretical or experimental study on the potential benefits of birefringent high power adiabatic second harmonic generation in terms of bandwidth and efficiency. In this paper, we aim to fill this information gap by studying the acceptance bandwidth, the required thermal profile and the dependence on pump power of adiabatic second harmonic generation in LBO.

\begin{figure}[htbp]
\centering
\fbox{\includegraphics[width=\linewidth]{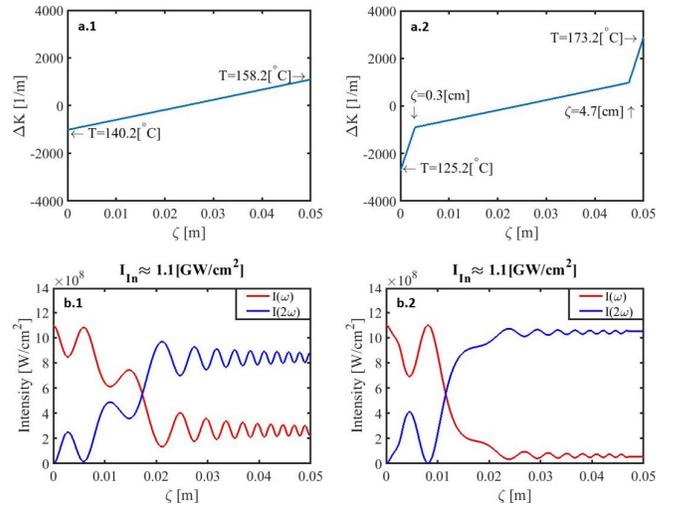}}
\caption{Comparison between linear phase mismatch profile (a.1) and apodized phase mismatch profile (a.2), for SHG in LBO crystal. For both cases, the temperature gradient in the intermediate region is the equal. (b.1)  and (b.2) corresponding variation in the pump and second harmonic intensities.}
\label{fig:picture2}
\end{figure}

The faster temperature gradients, near the entrance and exit of the crystal (shown in Fig. \ref{fig:picture1}a.2), were chosen to handle the inherent trade-off of the adiabatic process \cite{Porat:13}. The system should ideally start and finish near a stationary state in which the nonlinear coupling is negligible, i.e. $\Delta$k should satisfy \cite{Porat:13} |$\Delta$k|>>$\sqrt{P_3\kappa_1^2\kappa_3}$ and sign($\Delta$k$_{start}$)=-sign($\Delta$k$_{end}$). P$_3$ is proportional to the total photon flux, which is equal at the crystal entrance to the pump (fundamental only) flux and $\kappa_1$ ($\kappa_3$) are coupling coefficients from fundamental to the second harmonic (second harmonic to fundamental). To satisfy this, it is desired to have low values for the pump power and coupling. We emphasize that the eigenstate at the crystal's exit is very different from the one it started with. For the second harmonic process; in the initial state, all the light is at the fundamental frequency, whereas at the final state all the light is at the second harmonic. On the other hand, the stationary state needs to change very slowly, by changing the phase mismatch, in order to remain an eigenstate of the evolving system as it passes through the phase matching temperature. For the special case of adiabatic SHG, the normalized rate of change in the mismatch parameter at the intermediate region, expressed by the parameter $r_{nl}$ \cite{Porat:13}, must satisfy $r_{nl}=\dfrac{2}{\sqrt{27}}\dfrac{1}{P_3}\left|\dfrac{d(\Delta k/\sqrt{\kappa_1^2\kappa_3})}{d(\zeta\sqrt{\kappa_1^2\kappa_3})}\right|$<<1, $\zeta$ represents the propagation coordinate. This necessitates to have a large pump intensities and large coupling.
If we use a thermal profile based on a linear slope, it becomes difficult to maintain the necessary conditions. For example, if we choose a slope that satisfies the $r_{nl}$ criterion, as illustrated in Fig. \ref{fig:picture2}a.1, the temperatures at the two crystal edges are not very far from the phase matching temperature, and therefore when we inject only the pump wave the system is not at a stationary state since the pump and second harmonic are not decoupled. The conversion process is therefore oscillatory and does not provide high conversion efficiency, as shown in Fig. \ref{fig:picture2}b.1. On the other hand, it is not possible to maintain the $r_{nl}$ condition if we set the temperatures at the two crystal edges to be far from the phase matching temperature while setting a linear temperature slope.

We can overcome this problem by using an apodized phase-mismatch profile \cite{doi:10.1002/lpor.201300107,Phillips:13,Heese:12_}. This may be achieved by having higher temperature gradient near the entrance and exit of the crystal to satisfy decoupling of the pump and second harmonic waves, and a lower gradient in the intermediate part of the crystal to fulfill the adiabatic-slow variation of the moment-mismatch, as shown in Fig. \ref{fig:picture2}a.2. The unwanted intensity oscillations between the pump and second harmonic are suppressed and the conversion process becomes stable and more efficient, as shown in Fig. \ref{fig:picture2}b.2. For this profile, we have at the edges a phase mismatch $\Delta$k that is about 7 times larger than $\sqrt{P_3\kappa_1^2\kappa_3}$, and a normalized slope in the middle part of $r_{nl}\simeq0.09$.

\begin{figure}[htbp]
\centering
\fbox{\includegraphics[width=\linewidth]{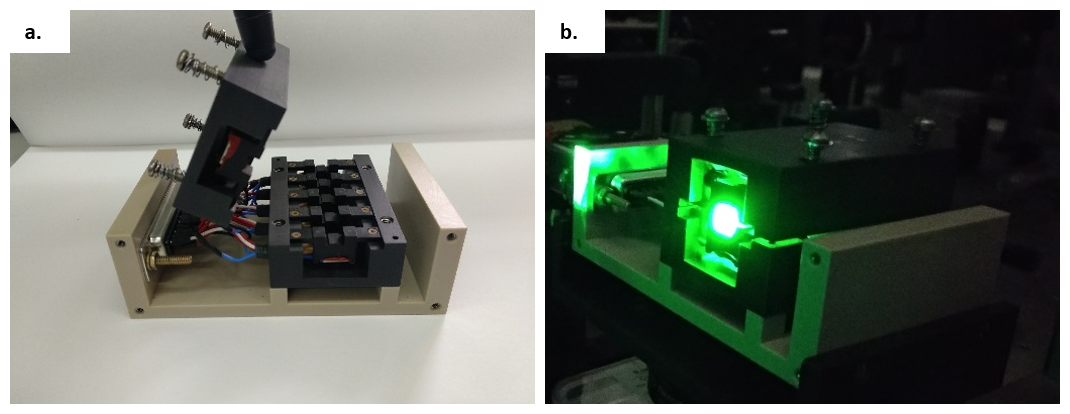}}
\caption{(a) Crystal oven for realizing the thermal gradients, with 5 temperature set points. (b) Adiabatic second harmonic generation in the LBO crystal. (the oven is shown without its cover).
}
\label{fig:picture3}
\end{figure}

\begin{figure}[htbp]
\centering
\fbox{\includegraphics[width=\linewidth]{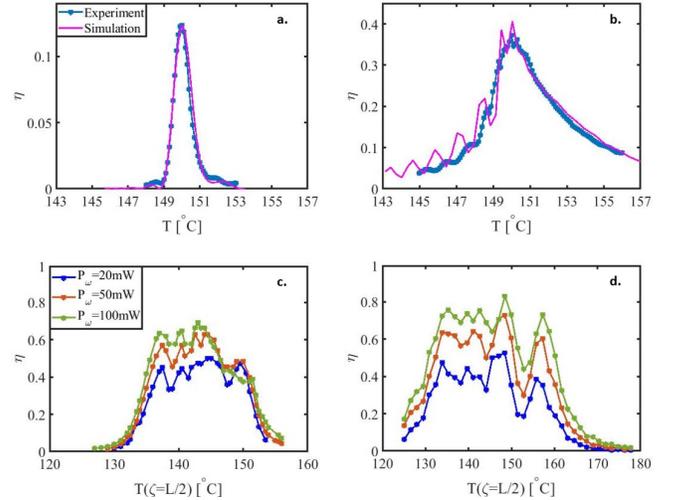}}
\caption{SHG conversion efficiency dependence on crystal temperature (a) for constant crystal temperature, undepleted pump. (b) for constant crystal temperature, depleted pump. Experimental (blue-*) and simulation (magenta-) results are shown.
Experimental results (c) and corresponding simulation (d) for SHG under thermal adiabatic profile, as a function of mean crystal temperature, for three input pump intensities.
}
\label{fig:picture4}
\end{figure}

A 5cm-long, 5 mm-thick non-critically phase matched LBO crystal, was selected as our experimental interaction medium. It is mounted in a temperature-controlled oven to maintain a type-I non-critically phase-matched SHG. To achieve the desired temperature profile, we have designed temperature-controlled oven (shown in Fig. \ref{fig:picture3}) with 5 independent set points located at the following distances from the crystal entrance plane – [0, 12.5, 25, 37.5, 50] mm, and the corresponding temperatures in them were set to – [Tm-14, Tm-5, Tm, Tm+5, Tm+14] $^{\circ}$C. This temperature profile was chosen to optimize the conversion efficiency and bandwidth, under the physical constraints set by the location of the five different heaters in the oven. If we choose Tm to be 149$^{\circ}$C ,perfect phase matching is obtained at the middle of the crystal, but we can also use other values of Tm to determine the thermal acceptance of the frequency doubler. The oven is designed such that the outer surfaces of the crystal are isolated from the environment by an air gap and a cover to minimize the heat exchange with its surroundings. The pump sources was a Q-Switched Nd:YAG pulsed laser producing 1.1 ns pulses with 100 Hz repetition rate at a wavelength of 1064 nm, having an average power up to 100 mW (or 0.91 mJ per pulse). The pump beam was focused at the center of the LBO crystal, with a waist of 130 $\mu$m, thus providing a maximum (initial, unconverted) peak intensity of $\sim$1.71 GW/cm$^2$. The pump polarization is $\hat{z}$ polarized and the generated field is $\hat{y}$ polarized. We note that the absorption of the fundamental and SH waves in the LBO crystal is negligible \cite{nikogos}. We assume that the transverse temperature-gradient is negligible, owing to the symmetric thermal boundary conditions, i.e. in each one of the five heating points, a metallic frame is surrounding the crystal from all four facets.
 To experimentally characterize the nonlinear process, we measured the thermal acceptance at a fixed pump power; we first measured the conversion efficiency as a function of the crystal temperature at low pump intensity of $\sim0.018$ GW/cm$^2$ for the case in which the crystal temperature is identical throughout the entire crystal, as shown in Fig. \ref{fig:picture4}a. To minimize pump depletion we used in this measurement a relatively low power Nd:YAG laser, with an average power of 25 mW, a pulse duration of 5 ns, a repetition rate of 10 KHz (hence  pulse energy of 2.5 microJoule) and a beam waist of $\sim30$ $\mu$m. The efficiency profile exhibits a narrow peak with a bandwidth of 0.95$^{\circ}$C. Since we use Gaussian beams, the measured curve is slightly different than the familiar sinc$^2$ curve that is obtained in the case of a plane wave. Measurement is in very good agreement with the theoretical calculation (all simulations take diffraction-effects into account). The pump laser was then replaced to a microchip laser, the pump power was increased to 50 mW (0.46 mJ), and we re-measured the conversion efficiency vs temperature, as shown in Fig. \ref{fig:picture4}b. The thermal bandwidth increases to $\delta$T$\simeq$3.3$^{\circ}$C, with a peak conversion efficiency of roughly 37\%. The high peak power enables efficient conversion at non-phase matched temperatures, resulting in a broader thermal bandwidth compared to the undepleted case.

Whereas these two measurements serve as references for the performance of the system in the non-adiabatic case, in Fig. \ref{fig:picture4}c we show measurements of the thermal acceptance in the case of adiabatic conversion. The thermal response is observed by shifting simultaneously the 5 temperature set points by identical values. The measurement is performed at 3 different pump power of 20, 50 and 100 mW (peak energy of 0.18, 0.46, 0.91 mJ). The conversion efficiency in these three cases exhibits a relatively wide range in which the peak conversion efficiency is nearly flat. As the input intensity increases, the maximum efficiency increases, while bandwidth roughly stays the same. The maximum efficiency is $\eta_{20 mW}\simeq50\%$, $\eta_{50 mW}\simeq63\%$, $\eta_{100 mW}\simeq70\%$, and the conversion thermal bandwidth is roughly $\delta$T$_{20 mW}$=16$^{\circ}$C, $\delta$T$_{50 mW}$=17$^{\circ}$C, $\delta$T$_{100 mW}$=18$^{\circ}$C. With the same power level of 50 mW that was used in \ref{fig:picture4}b, the thermal acceptance is considerably larger than what is obtained in the non-adiabatic case, and moreover, the efficiency of the adiabatic case is higher. The higher efficiency is a manifestation of the robustness of the adiabatic process to the relatively large angular spread of the depleted pump beam. The measurements are in good agreement with theoretical simulations, shown in \ref{fig:picture4}d. The simulation results show the same trend as our experimental results, but they do differ in the values of the bandwidth and conversion efficiencies. The difference may be partly explained by the temporal profile of the pulse; in the simulation, we consider a rectangular pulse with constant peak power, whereas the experimental pulse has a Gaussian-like shape. Our laser operates at a fixed wavelength, so we cannot measure directly the spectral acceptance. However, the measured thermal bandwidth of 18$^{\circ}$C  at 100 mW corresponds to a spectral bandwidth of 14 nm near 1064 nm.

\begin{figure}[htbp]
\centering
\fbox{\includegraphics[width=\linewidth]{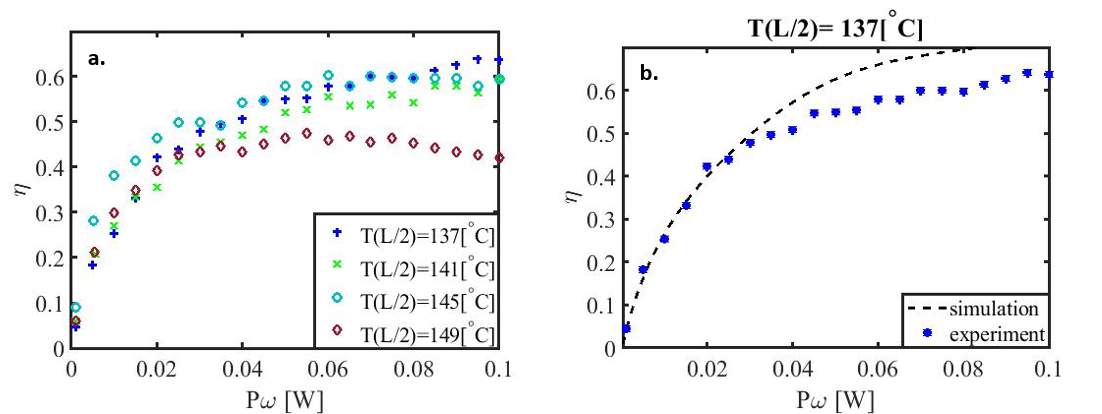}}
\caption{(a) Experimental results of adiabatic SHG efficiency as a function pump power, for 4 values of central temperature. (b) Simulation for a central temperature of 137$^{\circ}$C.
}
\label{fig:picture5}
\end{figure}

A valuable trait of the conversion process is its robustness under varying pump power. We expect that the efficiency of the process will not change significantly for a wide range of pump power levels. We measured the conversion efficiency vs pump power, as shown in Fig. \ref{fig:picture5}a, at four different central crystal temperatures Tm, but having the same thermal profile, shifted according to the central temperature. As expected for adiabatic interaction, the results indicate that conversion efficiency grows with input power. We obtained good agreement between the experiments and simulations, while the differences between the two may be attributed to the temporal shape of the pulse, as mentioned above. We may see that at low pump power, the hardest condition to satisfy is the adiabatic $r_{nl}$ condition, which relaxes as the intensity grows. At higher intensities, the requirement to coincide with eigenstates is harder to satisfy with the same temperature profile, so the efficiency saturates. We also note that the process is expected to be robust against variations of the crystal length and the angle of incidence.

An important conclusion is that the adiabatic properties are satisfied over a wide range of mean temperatures, implying its robustness against changes in various parameters. An interesting observation we can draw from Fig. \ref{fig:picture4}  and Fig. \ref{fig:picture5} is that for our experimental conditions, this process is satisfied below and up to the phase-matching-temperature of 149$^{\circ}$C  at the center of the crystal, but is not satisfied for higher temperatures. The explanation is that at lower temperatures the starting point is far enough from the phase matching temperature, so this ensures that the initial condition of pump wave only is an eigenstate of the system, and the system will, therefore, be maintained near an eigenstate till the exit surface. Ideally, the phase mismatch at that surface should be large enough so that the pump and second harmonic are again decoupled, hence all the light will be at the second harmonic wave. However if the phase mismatch is not large enough, this will only result in a reduced-efficiency. On the other hand, when the crystal is held at higher temperatures, the situation is completely different. In this case at the entrance facet, the pump and second harmonic wave are still coupled, so injecting only a pump wave means that right from the start the system is not at an eigenmode, hence it will not undergo an adiabatic transition to the second harmonic.

To summarize, a robust and high-intensity adiabatic second harmonic generation process was demonstrated, based on birefringent phase matching in LBO. This approach demonstrates the ability to use adiabatic processes at high power owing to the high damage threshold and a relatively large aperture of the crystal. The thermal gradient method that we used is easy to implement, and moreover, the thermal profile can be reconfigured for different pumping parameters. The method we presented may be advantageous in particular for frequency conversion of fiber lasers. These lasers have a wider spectral bandwidth compared to solid-state lasers, owing to nonlinear effects in the fiber \cite{Lapointe}, and significant efforts are made to maintain a narrow linewidth for nonlinear optical applications. We anticipate that the increased bandwidth of adiabatic frequency conversion processes may enable to save the efforts of maintaining narrow linewidth without sacrificing the conversion efficiency.

\subsection{Acknowledgement:}
This work was supported by the Israeli Innovation Authority and by the Israeli Science Foundation, grant no 1415/17.

\bigskip
\bibliography{references}


\end{document}